\documentclass[12pt,letterpaper]{article}

\usepackage{cite}
\usepackage{graphicx}
\usepackage{hyperref} %%% note: automatically loads "color", use "dvips -z"

\begin{document}

%%%%%%%%%%%%%%%%%%%%%%%%%%%%%%%%%%%%%%%%%%%%%%%%%%%%%%%%%%%%%%%%%%%%%%%%%%%%%%%

%\phantom{nothing}
\begin{center}
{\LARGE\bf Physics of $\eta'$ with rooted staggered quarks}
\end{center}
\vspace{10pt}

\begin{center}
{\large\bf Stephan D\"urr}
\\[6pt]
{\sl
Bergische Universit\"at Wuppertal,
Gau\ss$\,\!$stra\ss$\,\!$e\,20,
42119 Wuppertal, Germany
\\
J\"ulich Supercomputing Center,
Forschungszentrum J\"ulich,
52425 J\"ulich, Germany
}
\end{center}
\vspace{10pt}

\begin{abstract}
\noindent
%%%
%The quark-mass dependence of the $\eta$ in the Schwinger model, which --~like
%the $\eta'$ in QCD~-- receives part of its mass through the axial anomaly, is
%studied on the lattice for $N_f\!=\!0,1,2$ dynamical flavors. These are
%represented by staggered quarks, with a rooted determinant for $N_f\!=\!1$.
%%%
%The massive Schwinger model with $N_f\!=\!0,1,2$ is used to study the quark
%mass dependence of the $\eta$ which, like the $\eta'$ in QCD, obtains most of
%its mass through the axial anomaly. The flavors are represented by staggered
%quarks, with a rooted determinant for $N_f\!=\!1$.
%%%
%The quark-mass dependence of the $\eta$ in the Schwinger model, which --~like
%the $\eta'$ in QCD~-- receives a mass through the axial anomaly, is studied
%with $N_f\!=\!0,1,2$ dynamical flavors. The latter are represented by staggered
%quarks, with a rooted determinant for $N_f\!=\!1$. The results strongly suggest
%that for positive quark mass the rooting recipe is correct.
%%%
The quark-mass dependence of the $\eta$ in the Schwinger model, which --~like
the $\eta'$ in QCD~-- becomes massive through the axial anomaly, is studied
on the lattice with $N_{\!f}\!=\!0,1,2$. Staggered quarks are used, with a
rooted determinant for $N_{\!f}\!=\!1$. In the chiral limit the Schwinger mass
is reproduced, which suggests that the anomaly is being treated correctly.
%%%
%\hfill(\today)%begin:1.8.09
\end{abstract}
\vspace{4pt}

%%%%%%%%%%%%%%%%%%%%%%%%%%%%%%%%%%%%%%%%%%%%%%%%%%%%%%%%%%%%%%%%%%%%%%%%%%%%%%%

\newcommand{\pad}{\partial}
\newcommand{\hqu}{\hbar}
\newcommand{\ovr}{\over}
\newcommand{\til}{\tilde}
\newcommand{\pri}{^\prime}
\renewcommand{\dag}{^\dagger}
\newcommand{\<}{\langle}
\renewcommand{\>}{\rangle}
\newcommand{\gaf}{\gamma_5}
\newcommand{\lap}{\triangle}
\newcommand{\dal}{{\sqcap\!\!\!\!\sqcup}}
\newcommand{\trc}{\mathrm{tr}}
\newcommand{\Trc}{\mathrm{Tr}}
\newcommand{\Mpi}{M_\pi}
\newcommand{\Fpi}{F_\pi}
\newcommand{\Mka}{M_K}
\newcommand{\Fka}{F_K}
\newcommand{\Met}{M_\et}
\newcommand{\Fet}{F_\et}

\newcommand{\al}{\alpha}
\newcommand{\be}{\beta}
\newcommand{\ga}{\gamma}
\newcommand{\de}{\delta}
\newcommand{\ep}{\epsilon}
\newcommand{\ve}{\varepsilon}
\newcommand{\ze}{\zeta}
\newcommand{\et}{\eta}
\renewcommand{\th}{\theta}
\newcommand{\vt}{\vartheta}
\newcommand{\io}{\iota}
\newcommand{\ka}{\kappa}
\newcommand{\la}{\lambda}
\newcommand{\rh}{\rho}
\newcommand{\vr}{\varrho}
\newcommand{\si}{\sigma}
\newcommand{\ta}{\tau}
\newcommand{\ph}{\phi}
\newcommand{\vp}{\varphi}
\newcommand{\ch}{\chi}
\newcommand{\ps}{\psi}
\newcommand{\om}{\omega}

\newcommand{\psb}{\bar{\psi}}
\newcommand{\etb}{\bar{\eta}}
\newcommand{\psh}{\hat{\psi}}
\newcommand{\eth}{\hat{\eta}}
\newcommand{\psd}{\psi^{\dagger}}
\newcommand{\etd}{\eta^{\dagger}}
\newcommand{\qh}{\hat{q}}
\newcommand{\kh}{\hat{k}}

\newcommand{\bdm}{\begin{displaymath}}
\newcommand{\edm}{\end{displaymath}}
\newcommand{\bea}{\begin{eqnarray}}
\newcommand{\eea}{\end{eqnarray}}
\newcommand{\beq}{\begin{equation}}
\newcommand{\eeq}{\end{equation}}

\newcommand{\mr}{\mathrm}
\newcommand{\mb}{\mathbf}
\newcommand{\Nf}{N_{\!f\,}}%{{N_{ f }}}
\newcommand{\Nc}{N_{  c  }}%{{N_{ c }}}
\newcommand{\Nt}{N_{  t  }}%{{N_{ t }}}
\newcommand{\ri}{\mr{i}}
\newcommand{\DW}{D_\mr{W}}
\newcommand{\DN}{D_\mr{N}}
\newcommand{\Dov}{D_\mr{ov}}
\newcommand{\Dst}{D_\mr{st}}
\newcommand{\Dmi}{D_\mr{mi}}
\newcommand{\Dovm}{D_{\mr{ov},m}}
\newcommand{\Dstm}{D_{\mr{st},m}}
\newcommand{\Dmim}{D_{\mr{mi},m}}
\newcommand{\MeV}{\,\mr{MeV}}
\newcommand{\GeV}{\,\mr{GeV}}
\newcommand{\fm}{\,\mr{fm}}
\newcommand{\MSbar}{\overline{\mr{MS}}}
\newcommand{\nab}{\nabla}

\long\def\begincomment#1\endcomment{}

\hyphenation{topo-lo-gi-cal simu-la-tion theo-re-ti-cal mini-mum con-tinu-um}

%%%%%%%%%%%%%%%%%%%%%%%%%%%%%%%%%%%%%%%%%%%%%%%%%%%%%%%%%%%%%%%%%%%%%%%%%%%%%%%

\section{Introduction}

%%%%%%%%%%%%%%%%%%%%%%%%%%%%%%%%%%%%%%%%%%%%%%%%%%%%%%%%%%%%%%%%%%%%%%%%%%%%%%%

Staggered fermions \cite{Susskind:1976jm} offer a cost-effective way of
regulating QCD with four degenerate species.
In nature, the four lightest quarks, known as the $u,d,s,c$ flavors, are far
from being degenerate; only the $u$ and $d$ quarks are approximately degenerate
in the sense that $m_d\!-\!m_u\!\ll\!\Lambda_\mr{had}$, where
$\Lambda_\mr{had}\!=\!O(1\GeV)$ denotes a typical hadronic scale.
Since for an integer number of degenerate dynamical flavors the functional
measure of QCD scales with the $\Nf$-th power of the determinant, it has been
proposed \cite{Marinari:1981qf} to reverse this relationship, and to represent
$\Nf\!=\!2\!+\!1$ QCD at finite lattice spacing $a$ (or cut-off $a^{-1}$) by
the Euclidean partition function
\beq
Z=\int\!DU\;
\det\nolimits^{1/2}(D_{\mr{stag},m_{ud}})\,
\det\nolimits^{1/4}(D_{\mr{stag},m_s})\,e^{-S_\mr{g}}
\label{def_fourthroot}
\eeq
where the path-integral runs over all gauge backgrounds $U$, and $S_\mr{g}$
denotes the gauge action.
Thus the square-root of the determinant of a staggered field with the isospin
averaged light quark mass $m_{ud}\!=\!(m_u\!+\!m_d)/2$ and the fourth-root of
the determinant of a field with the strange quark mass $m_s$ are utilized to
define the regulated version of QCD which is used in several state-of-the-art
studies of phenomenologically relevant quantities (see e.g.\
\cite{Bazavov:2009bb}).

In recent years, the setup (\ref{def_fourthroot}) has been criticized
\cite{Jansen:2003nt,Creutz:2007yg,Creutz:2007rk}, because there is no
field-theoretic proof that its continuum limit is really QCD, or put
differently, that the lattice theory (\ref{def_fourthroot}) is in the correct%
\footnote{In addition to the summary talks \cite{Durr:2005ax,Sharpe:2006re,
Kronfeld:2007ek,Golterman:2008gt}, the interested reader is referred to the
Schwinger model condensate tests of \cite{Azcoiti:1994av,Durr:2003xs,
Durr:2004ta}, the eigenvalue based arguments of \cite{Durr:2003xs,
Follana:2004sz,Durr:2004as,Follana:2005km,Adams:2009eb}, the analysis in rooted
staggered chiral perturbation theory \cite{Sharpe:2004is,Bernard:2006zw}, the
renormalization-group based arguments of \cite{Bernard:2006ee,Shamir:2006nj},
and the analysis of the 't\,Hooft vertex \cite{Donald:2011if}.}
universality class.
The issue is more involved than (\ref{def_fourthroot}) would suggest, since in
practice one needs the generating functional $Z[\bar\et,\et]$ rather than the
partition function, and the manner in which a given staggered field is reduced
to a single ``taste'' (the modern word for a single species within a staggered
field) in the valence sector differs from the rooting recipe
(\ref{def_fourthroot}) that is applied in the sea sector of the theory.
Accordingly, the question is whether these two reduction mechanisms work in
concert, to define a valid discretization of QCD.

As much of the criticism focuses on the axial anomaly and the special role
played by the $\et'$ in QCD \cite{Creutz:2007yg,Creutz:2007rk}, a detailed
investigation of this state seems particularly desirable.
The $\et'$ requires disconnected contributions, and this poses a technical
challenge.
However, since the underlying physics is common to a broad class of vector-like
gauge theories, there is no need to attack the problem in QCD.
In this article the flavor-singlet state is studied in the generalized
Schwinger model (QED in 2D, with $\Nf\!=\!0,1,2$ massive flavors)
\cite{Schwinger:1962tp}, which is much easier to simulate.
In 2D a staggered field contains only two species.
Hence, a square-root is required for $\Nf\!=\!1$, while the $\Nf\!=\!2$
continuum limit is supposed to be correct by definition.
The point is that the conceptual issues match those of QCD.
The $\et$ in this model plays the same role as the $\et'$ in QCD with three
dynamical flavors, since its mass is predominantly due to the (global) axial
anomaly.
In the chiral limit of the $\Nf\!=\!1$ theory it is known as the Schwinger
particle.

%%%%%%%%%%%%%%%%%%%%%%%%%%%%%%%%%%%%%%%%%%%%%%%%%%%%%%%%%%%%%%%%%%%%%%%%%%%%%%%

\section{Simulation setup}

%%%%%%%%%%%%%%%%%%%%%%%%%%%%%%%%%%%%%%%%%%%%%%%%%%%%%%%%%%%%%%%%%%%%%%%%%%%%%%%

The goal is to perform $\et$ spectroscopy with (rooted) staggered quarks in the
massive Schwinger model ($\Nf\!=\!0,1,2$) at several values of the coupling,
such that the continuum limit ($a\!\to\!0$) can be taken.
In the $\Nf\!=\!1$ case we wish to perform, in the second step, a chiral
extrapolation to compare to the analytic prediction $\Met^2\!=\!e^2/\pi$ at
$m\!=\!0$ by Schwinger \cite{Schwinger:1962tp}.

%\begin{table}[!tb]
%\centering
%\begin{tabular}{|cc|cc|cc|r|}
%\hline
% 1.8 & $12^2$ &  1.6 & $16^2$ &  1.8 & $24^2$ & 80'000 \\
% 3.2 & $16^2$ &  3.6 & $24^2$ &  3.2 & $32^2$ & 40'000 \\
% 7.2 & $24^2$ &  6.4 & $32^2$ &  7.2 & $48^2$ & 20'000 \\
%12.8 & $32^2$ & 14.4 & $48^2$ & 12.8 & $64^2$ & 10'000 \\
%\hline
%\end{tabular}
%\caption{\sl
%Lattices with $N^2/\be$ equal to 80 (left), 160 (center), 320 (right).}
%\end{table}

\begin{table}[!tb]
%%% beta:=vector([1.8,3.2,7.2,12.8,3.2,3.2]);
%%% L:=vector([24,32,48,64,24,40]);
%%% m:=vector([0.032,0.024,0.016,0.012,0.024,0.024]);
%%% for i from 1 to 6 do
%%%    print(L[i]/sqrt(beta[i]),\
%%%     sqrt(1.)/sqrt(evalf(Pi)*beta[i]),\
%%%     sqrt(2.)*L[i]/sqrt(evalf(Pi)*beta[i]),\
%%%     L[i]*(2.008^3*(m[i]*1)^2/sqrt(beta[i]))^(1./3.),\
%%%     L[i]*(2.008^3*(m[i]*2)^2/sqrt(beta[i]))^(1./3.),\
%%%     L[i]*(2.008^3*(m[i]*3)^2/sqrt(beta[i]))^(1./3.),\
%%%     L[i]*(2.008^3*(m[i]*4)^2/sqrt(beta[i]))^(1./3.),\
%%%     L[i]*(2.008^3*(m[i]*5)^2/sqrt(beta[i]))^(1./3.))
%%% end do;
\centering
\begin{tabular}{|cc|ccc|c|}
\hline
$\be$ & $L/a$ & $am$ & $\Met L$ & $\Mpi L$ & \#confs \\
\hline
 1.8&24&$0.032\cdot\!\{1,2,3,4,5\}$&14.27&$\{4.40,6.99,9.16,11.10,12.88\}$&$5\!\cdot\!80\,000$ \\
 3.2&32&$0.024\cdot\!\{1,2,3,4,5\}$&14.27&$\{4.40,6.99,9.16,11.10,12.88\}$&$5\!\cdot\!40\,000$ \\
 7.2&48&$0.016\cdot\!\{1,2,3,4,5\}$&14.27&$\{4.40,6.99,9.16,11.10,12.88\}$&$5\!\cdot\!20\,000$ \\
12.8&64&$0.012\cdot\!\{1,2,3,4,5\}$&14.27&$\{4.40,6.99,9.16,11.10,12.88\}$&$5\!\cdot\!10\,000$ \\
\hline
 3.2&24&$0.024\cdot\!\{1,2,3,4,5\}$&10.70&$\{3.30,\;5.24,\;6.87,\;8.32,\;9.66\}$&$5\!\cdot\!40\,000$ \\
 3.2&40&$0.024\cdot\!\{1,2,3,4,5\}$&17.84&$\{5.51,  8.74, 11.45, 13.87, 16.10\}$&$5\!\cdot\!40\,000$ \\
\hline
\end{tabular}
\caption{\sl\label{tab:setup}
Overview of the $eL\!=\!17.89$ simulations (top). The entries in the $\Met L$
(for $m\!=\!0)$ and $\Mpi L$ columns quote the predictions for $\Nf\!=\!2$
\cite{Schwinger:1962tp,Smilga:1996pi}. For $\Nf\!=\!1$ only the former exists
and is smaller by a factor $\sqrt{2}$. To test for finite volume effects, the
$\be\!=\!3.2$ runs are repeated at $eL\!=\!13.42,22.36$ (bottom). Simulations
are performed at $\Nf\!=\!0$, with reweighting to $\Nf\!=\!1,2$.}
\end{table}

Because of the super-renormalizability of the Schwinger model
\cite{Schwinger:1962tp}, a convenient choice of scale is through the
dimensionful coupling $e$ in $\be\!=\!1/(ae)^2$.
With this choice it is then straightforward to select the spatial extent
$L_1\!\equiv\!L$, the temporal extent $L_2\!\equiv\!T$ and the coupling $\be$
such that $eL$ is fixed (modulo cut-off effects).
Moreover, due to the predictions of the eta mass in the chiral limit
($\Met^2\!=\!\Nf e^2/\pi$ for $\Nf\!=\!1,2$, see \cite{Schwinger:1962tp}) and of
the pion mass as a function of the quark mass ($\Mpi\!=\!2.008e^{1/3}m^{2/3}$
for $\Nf\!=\!2$, see \cite{Smilga:1996pi}), one knows $\lim_{m\to0}\Met L$
beforehand, and one may choose the quark masses, at least for $\Nf\!=\!2$, such
that $\Mpi L$ assumes predefined values (again modulo cut-off effects).
The parameters of the square lattices used in this article are shown in
Tab.\,\ref{tab:setup}.
Most of them yield $eL\!=\!eT\!=\!17.89$, but at one coupling (with five quark
masses) dedicated finite-volume scaling studies are performed (see below).

The covariant derivative in $D_{\mr{stag},m}$ (and in the staggered 2-hop
operators described below) uses 3 steps of APE smearing
%$V_\mu(x)=P_{U(1)}\{U_\mu(x)\!+\!U_\nu(x)U_\mu(x\!+\!\hat\nu)
%U_\nu(x\!+\!\hat\mu)\dag\!+\!U_\nu(x\!-\!\hat\nu)\dag
%U_\mu(x\!-\!\hat\nu)U_\nu(x\!+\!\hat\mu-\!\hat\nu)$
with $\al\!=\!0.5$ and back-projection to U(1).
A great simplification is that reweighting techniques prove effective in 2D
\cite{Lang:1997ib}.
The plan is thus to generate substantial numbers of quenched lattices, and to
include the determinant factor into the observable.
Using standard LAPACK routines, $\log(\det(D_{\mr{stag},m}))$ may be calculated
via a Cholesky factorization $D_{\mr{stag},m}\dag D_{\mr{stag},m}\!=\!R\dag R$
by summing logs of the diagonal elements of $R$.
With $R$ in hand, $D_{\mr{stag},m}^{-1}$ may be computed through two solve-for
operations.
Alternatively, one may use the LU decomposition with pivoting to compute the
determinant and the inverse.

%%%%%%%%%%%%%%%%%%%%%%%%%%%%%%%%%%%%%%%%%%%%%%%%%%%%%%%%%%%%%%%%%%%%%%%%%%%%%%%

\section{Topological charge decorrelation}

%%%%%%%%%%%%%%%%%%%%%%%%%%%%%%%%%%%%%%%%%%%%%%%%%%%%%%%%%%%%%%%%%%%%%%%%%%%%%%%

Since the disconnected part of the $\et$ correlator (for $\Nf\!=\!1,2$) is
sensitive to the global topological charge $q$ of the gauge background $U$
\cite{Fukaya:2003ph,Bietenholz:2011ey}, it is crucial to achieve excellent
ergodicity w.r.t.\ $q$.

In the Schwinger model the electric flux of a gauge configuration is quantized;
with toroidal boundary conditions [and ignoring a subset of configurations with
measure zero] the quantity
\beq
q[U]={1\ovr4\pi}\int\!d^2x\;\ve_{\mu\nu}F_{\mu\nu}(x)\;=\;
{1\ovr2\pi}\int\!d^2x\;F_{12}(x)
\eeq
takes on integer values.
Field configurations with constant flux are known as instantons.
They minimize the action in a given charge sector, but unlike in QCD they
are completely delocalized.
Such one-instanton configurations can be directly put on the lattice.
Specifically \cite{Smit:1986fn}
\beq
\begin{array}{rcl}
U_1(x)&=&\exp(-2\pi\ri x_2/[L_1L_2])
\\[2mm]
U_2(x)&=&\exp(+2\pi\ri x_1/L_1\cdot\de_{x_2,L_2})
\end{array}
\label{def_smit}
\eeq
is an implementation in pseudo-Coulomb gauge (where all links in the time
direction are one, except those at $x_2\!=\!L_2$).
An anti-instanton follows by reversing the signs in the exponent.

\begin{figure}[!tb]
\includegraphics[height=67mm]{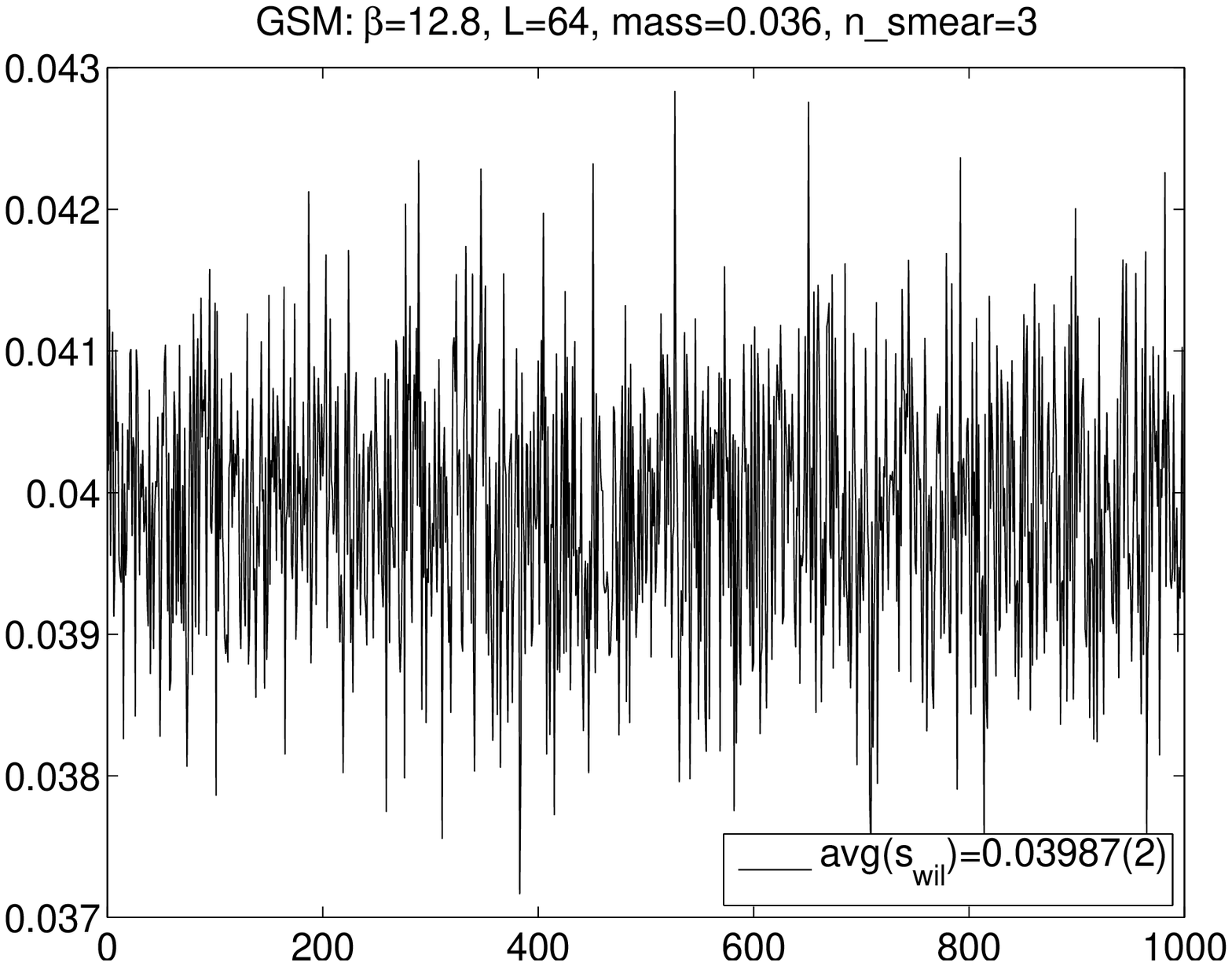}
\includegraphics[height=67mm]{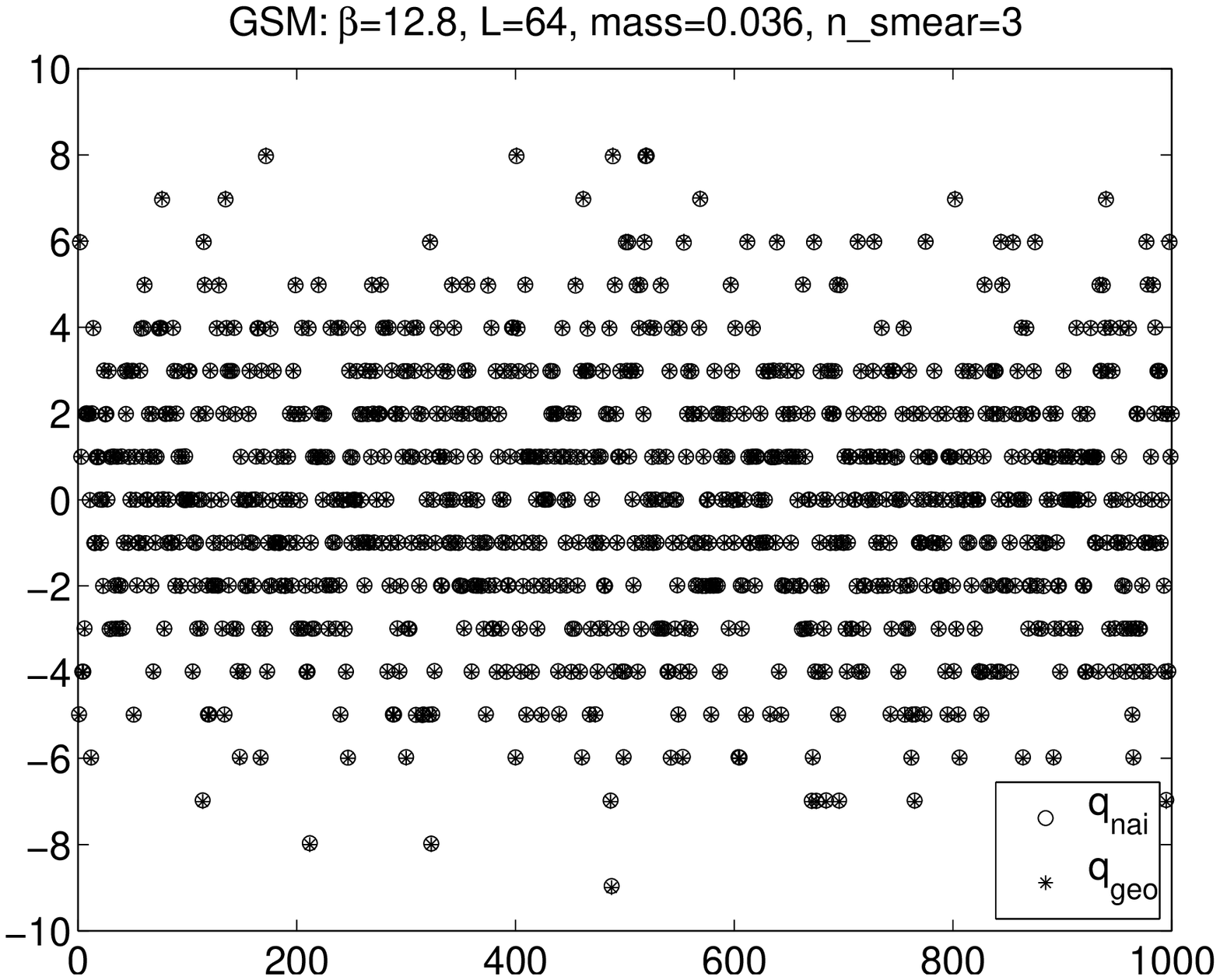}
\caption{\sl\label{fig:history}
Part of the $\Nf\!=\!0$ Monte Carlo histories of the action $s_\mr{wil}$ (left)
and of the 3-fold smeared field-theoretic topological charges
$q_\mr{nai}^{(3)},q_\mr{geo}^{(3)}$ (right) on the $64^2$ lattices at
$\be\!=\!12.8$.}
\end{figure}

It is thus natural to realize a topology-changing update by multiplying a given
configuration, link by link, with (\ref{def_smit}) or its conjugate (equal
probability), the result being subject to a Metropolis accept-reject step.
This way detailed balance is maintained by construction.
With dynamical fermions, one would calculate
$\det^{\Nf}(D[U_\mr{new}])/\det^{\Nf}(D[U_\mr{old}])$ and include it
into the decision.
Upon combining this global update with a standard procedure
\cite{Metropolis:1953am,Brown:1987rra}, one has a simulation algorithm of a
two-dimensional%
\footnote{In fact, this idea may be used in the U(1) theory in 4D too, by just
combining two such planar instantons (e.g.\ in the 12 and 34-planes, see
\cite{Smit:1986fn} for details), albeit with the proviso that $q$ then changes
only by $\pm2$ units.}
U(1) gauge theory which may take large steps%
\footnote{To avoid any correlation between adjacent configurations, on average
a total drift by $(\chi_\mr{top}V)^{1/2}\!=\!\<q^2\>^{1/2}$ units must
be realized, tantamount to a successful completion of approximately
$\chi_\mr{top}V\!=\!\<q^2\>$ changes of $q$.}
in configuration space.
To further improve the symmetry of the overall charge distribution, one may
perform, once in a while, a P-transformation of the gauge field, i.e.\ apply
the transformation \cite{Leinweber:2003sj}
\beq
U_\mu(\mb{x},t)\to U_\mu(-\mb{x}-\hat\mu,t)\dag\quad(\mu\!=\!1,..,d\!-\!1)
\;,\quad
U_\mu(\mb{x},t)\to U_\mu(-\mb{x},t)\quad(\mu\!=\!d)
\eeq
which leaves the gauge action invariant but reverses the topological charge.

In this paper, an $\Nf\!=\!0$ gauge update consists of 4 over-relaxation sweeps
\cite{Brown:1987rra} per Metropolis sweep \cite{Metropolis:1953am} (each
Metropolis update applies 4 successive hits on a given link), and all of this
is repeated $L/(4a)$ times.
In total, two adjacent configurations are separated as\\[+2pt]
\tt\footnotesize
\phantom{------}for n=1:L/4\\[-4pt]
\phantom{---------}perform 2 over-relaxation sweeps\\[-4pt]
\phantom{---------}perform 1 instanton/anti-instanton update\\[-4pt]
\phantom{---------}perform 2 over-relaxation sweeps\\[-4pt]
\phantom{---------}perform 1 Metropolis sweep with 4 hits per link\\[-4pt]
\phantom{------}end\\[-4pt]
\phantom{------}perform, with probability 0.5, a P-transformation\\[+6pt]
\rm\normalsize
and Fig.\,\ref{fig:history} shows a part of the pertinent Monte-Carlo history
of the plaquette action and of the topological charge on the finest lattice
($\be\!=\!12.8$).
The Wilson action per site is defined as
$s_\mr{wil}(x)\!=\!1\!-\!\mr{Re}\,U_{12}(x)\!=\!1\!-\!\cos(\th_{12})$ with
$U_{12}(x)\!=\!U_1(x)U_2(x\!+\!e_1)U_1\dag(x\!+\!e_2)U_2\dag(x)\!=\!
\exp(\ri\th_{12})$ denoting the plaquette.
For the topological charge two field-theoretic definitions%
\footnote{For $q_\mr{nai}^{(n)}$ normally a renormalization factor
$Z\!=\!1\!+\!O(a^2)$ is introduced, but with smearing this factor is so close
to 1 that it seems permissible to drop it [in line with neglecting other
$O(a^2)$ effects].}
are used, $q_\mr{nai}^{(3)}\!=\!\sum\;\sin(\th_{12}^{(3)})/(2\pi)\!\in\!\mb{R}$
(``naive'') and $q_\mr{geo}^{(3)}\!=\!\sum\;\th_{12}^{(3)}/(2\pi)\!\in\!\mb{Z}$
(``geometric''), where $\th_{12}^{(3)}$ denotes the plaquette angle after $3$
smearing steps.
The algorithm is seen to tunnel well.

%%%%%%%%%%%%%%%%%%%%%%%%%%%%%%%%%%%%%%%%%%%%%%%%%%%%%%%%%%%%%%%%%%%%%%%%%%%%%%%

\section{Staggered spectroscopy with all-to-all technology}

%%%%%%%%%%%%%%%%%%%%%%%%%%%%%%%%%%%%%%%%%%%%%%%%%%%%%%%%%%%%%%%%%%%%%%%%%%%%%%%

%%%
%Staggered fermions are sensitive to the axial anomaly, if the staggered tastes
%are constructed in momentum space \cite{Karsten:1980wd,Sharatchandra:1981si}.
%However, the alternative construction in position space
%\cite{Gliozzi:1982ib,Duncan:1982xe,KlubergStern:1983dg} is the one
%which is used in practice (see e.g.\ \cite{Bazavov:2009bb}).
%Therefore, it is important to demonstrate that this formulation/interpretation
%of staggered fermions treats the axial anomaly correctly.
%%%
%Staggered fermions are sensitive to the axial anomaly, if the staggered tastes
%are constructed in momentum space \cite{Karsten:1980wd,Sharatchandra:1981si}.
%However, this is not exactly identical to the taste identification that is
%used in practice (see e.g.\ \cite{Bazavov:2009bb} for details), and it is thus
%important to demonstrate that the latter formulation/interpretation of
%staggered fermions treats the axial anomaly correctly.
%%%
In weak coupling perturbation theory staggered fermions have been proven to be
sensitive to the axial anomaly \cite{Karsten:1980wd,Sharatchandra:1981si}.
Still, it seems worth demonstrating that this sensitivity carries over, at the
non-perturbative level, to the asymptotic states of the theory, and leads to a
non-vanishing mass of the combined taste-and-flavor singlet pseudoscalar meson
in the chiral limit.
%%%

The remnant staggered form of the continuum index theorem has been investigated
in a classic paper \cite{Smit:1986fn}.
A direct check of the mass excess of the $\et$ ($\et'$) in $\Nf\!=\!2$
($\Nf\!=\!3$) QCD with rooted staggered quarks and standard taste assignment
has been attempted \cite{Venkataraman:1997xi,Gregory:2007ev,
Gregory:2008mn,Gregory:2011sg}, but unfortunately in the disconnected
contributions the signal dies quickly in the noise \cite{Struckmann:2000bt,
Lesk:2002gd,Schilling:2004kg,Hashimoto:2008xg,Christ:2010dd,Ottnad:2011mp}.
Encouraged by \cite{Fukaya:2003ph,Bietenholz:2011ey}, we now attack the same
goal in the much simpler Schwinger model.

With a single staggered field, we expect to find 4 pseudoscalar bosons.
The lightest ($\gaf\!\otimes\!\xi_5$, to be dubbed $\pi^0$) becomes massless
(on an infinitely large lattice) in the limit $m\!\to\!0$, the next two
($\gaf\!\otimes\!\xi_1$ and $\gaf\!\otimes\!\xi_1\xi_5$, to be dubbed
$\pi^\pm$) become massless up to cut-off effects, while the last one
($\gaf\!\otimes\!1$, to be dubbed $\et$) is supposed to be well separated and
stay massive in the chiral limit.
In standard terminology, the $\pi^0$ is taste-pseudoscalar, the $\pi^\pm$
are taste-(axial)vector (in 2D there is no distinction), while the $\et$ is
taste-scalar (or taste-singlet) -- see \cite{Bazavov:2009bb} for details.

\begin{figure}[!tb]
\includegraphics[width=0.49\textwidth]{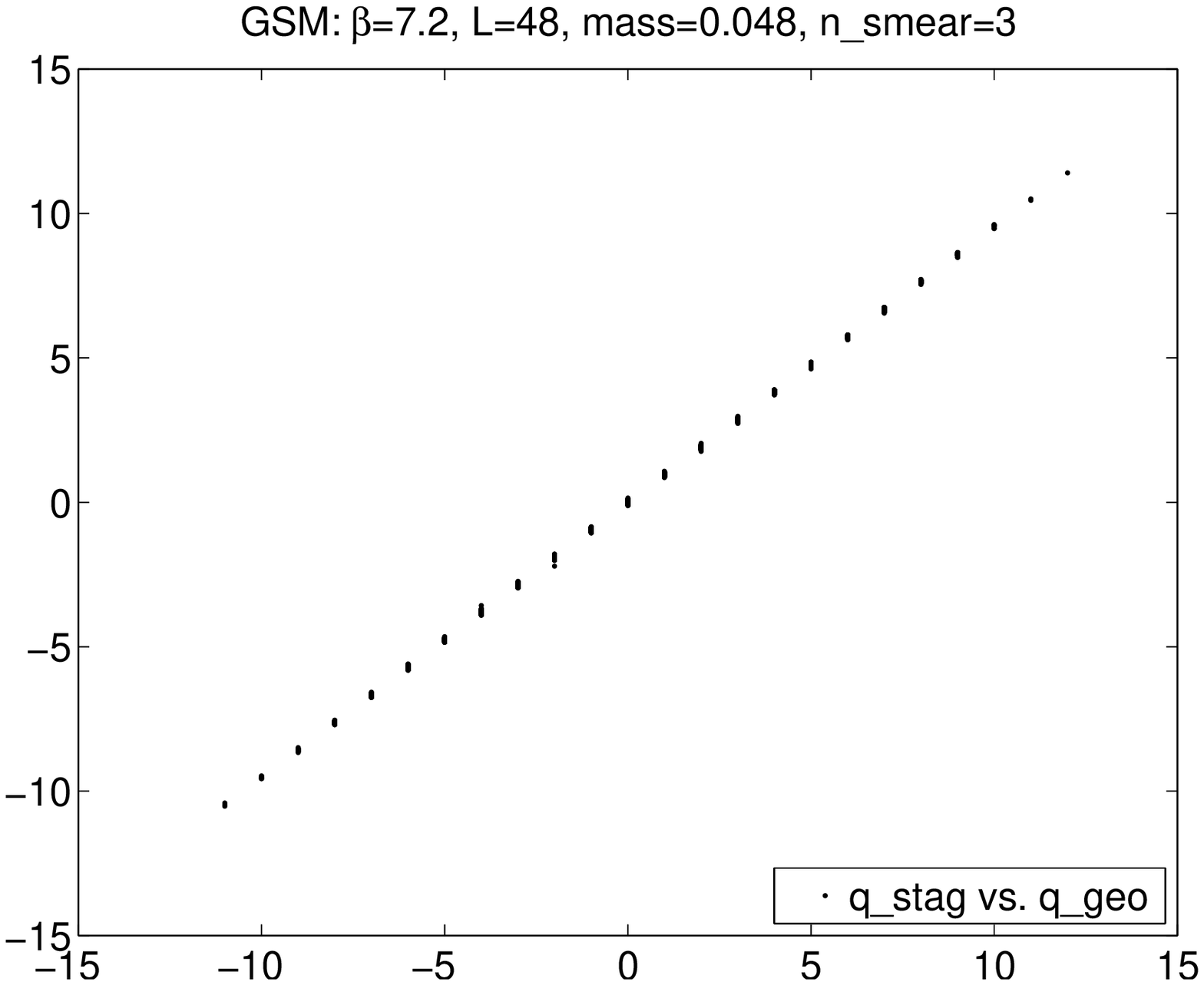}
\includegraphics[width=0.49\textwidth]{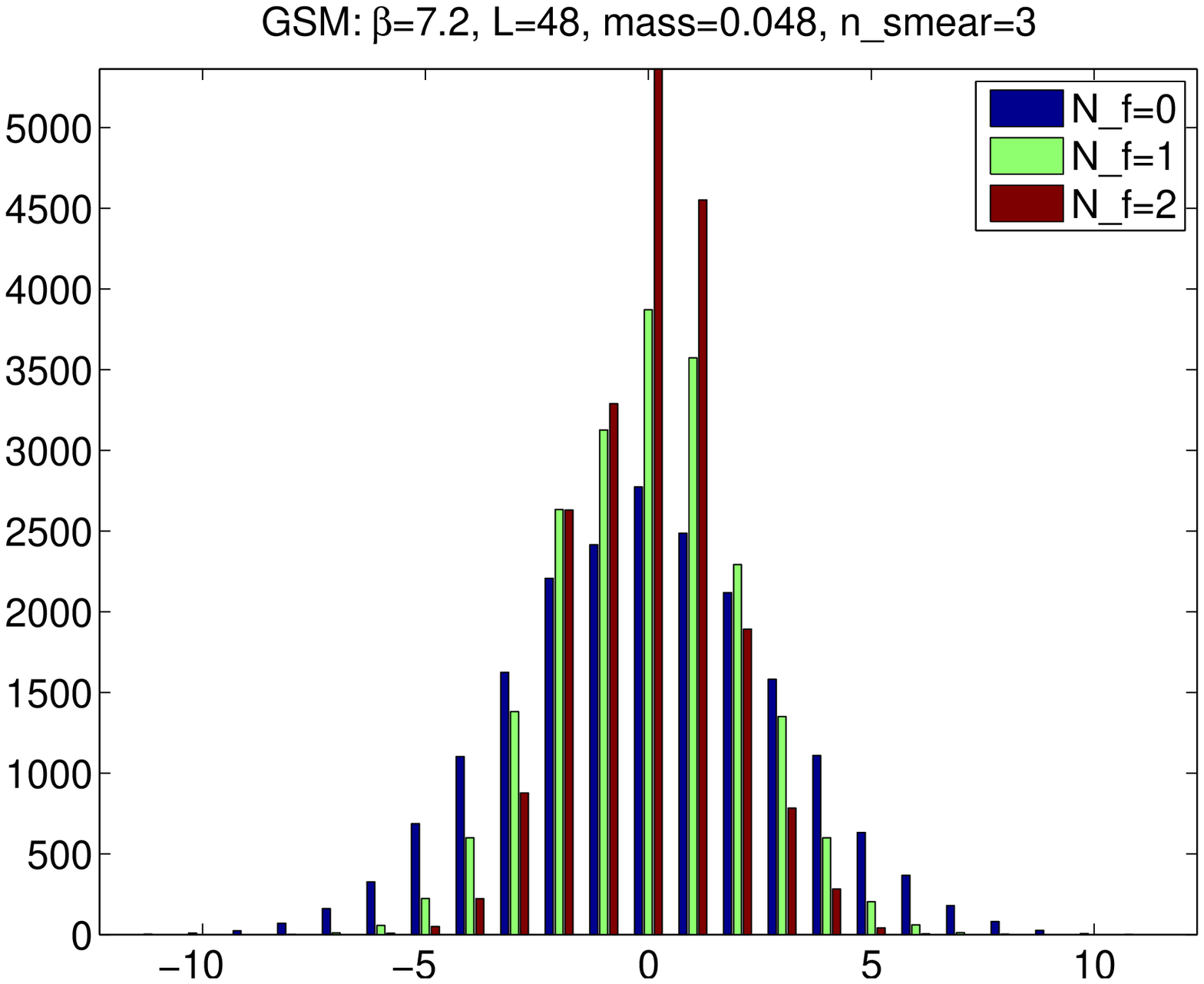}
\caption{\sl\label{fig:topology}
Left: Correlation between the topological charges $q_\mr{geo}^{(3)}$ and
$q_\mr{stag}^{(3)}$, based on the $\Gamma_{50}$ operator (\ref{def_50}), with 3
smearings. Right: Overall topological charge distribution for $\Nf\!=\!0$, and
after reweighting to $\Nf\!=\!1,2$. Either plot refers to the intermediate mass
run at $\be\!=\!7.2$.}
\end{figure}

Staggered spectroscopy is performed by constructing dedicated correlators which
project to a specific spinor-taste combination.
In phenomenological applications it is common practice to use a source at a
single lattice point as a ``broad band emitter'' which couples to all
spinor-taste combinations, and to apply the projection only at the sink.
In this work, we use an all-to-all propagator technique, and shall apply
spinor-taste projection independently at the source and the sink.
In explicit terms, for the $\gaf\!\otimes\!\xi_5$ ``Goldstone'' state we use
the operator $\Gamma_{55}=(-1)^{x_1+x_2}$, which is a point-like operator.
By contrast, the operator $\Gamma_{50}\simeq(\gaf\!\otimes\!1)$ is defined by
\cite{Smit:1986fn}
\bea
\Gamma_{50}&=&\frac{\ri}{2}(\Gamma_1\Gamma_2-\Gamma_2\Gamma_1)
\quad\mbox{where}
\label{def_50}
\\
\Gamma_\mu(x,y)&=&\frac{1}{2}\et_\mu(x)
\Big[U_\mu(x)\de_{x+\hat\mu,y}+U_\mu\dag(x\!-\!\hat\mu)\de_{x-\hat\mu,y}\Big]
\eea
with $\et_\mu(x)=(-1)^{\sum_{\nu<\mu}x_\nu}$, and is thus a 2-hop operator (it
would be 4-hop in 4D).

The operator (\ref{def_50}) is supposed to be sensitive to the topological
charge of the gauge background $U$, and to test our implementation we routinely
determine the fermionic charge%
\footnote{For $q_\mr{stag}^{(n)}$ the statement in footnote 3 applies likewise
(cf.\ Fig.\,\ref{fig:topology} to see how close to 1 the $Z$-factor is).}
\beq
q_\mr{stag}[U]=\frac{m}{2}\,\trc(G\Gamma_{50})
\label{q_stag}
\eeq
where $G$ denotes the Green's function of the massive staggered operator.
A scatter plot of (\ref{q_stag}) ($y$-axis) versus the gluonic charge
$q_\mr{geo}[U]$ ($x$-axis) is shown in Fig.\,\ref{fig:topology}.
We also checked that the operator $\Gamma_{50}\Gamma_{55}\simeq1\!\otimes\xi_5$
is not sensitive to the topological charge.

With these projections it is straightforward to construct the connected $\pi^0$
and $\et$ correlators
\bea
C_\pi(t)&=&\frac{1}{L^2}\sum_{x,y, x_2-y_2=t\,(\mr{mod}\,T)}
G(x,y)\Gamma_{55}(y) G(y,x)\Gamma_{55}(x)
\label{pgb_conn}
\\
C_\et(t)&=&\frac{1}{L^2}\sum_{x,y, x_2-y_2=t\,(\mr{mod}\,T)}
G(x,y')\Gamma_{50}(y',y) G(y,x')\Gamma_{50}(x',x)
\label{eta_conn}
\eea
where $\de(x',x)\Gamma_{55}(x)\!=\!\Gamma_{55}(x',x)$ is used, and the primed
positions are implicitly summed over.
For the combined taste-and-flavor singlet state there is also the disconnected
contribution
\beq
D_\et(t)\;=\;\frac{1}{L^2}\sum_{x,y, x_2-y_2=t\,(\mr{mod}\,T)}
G(x,x')\Gamma_{50}(x',x) G(y,y')\Gamma_{50}(y',y)
\label{eta_disc}
\eeq
and for staggered fermions it must be combined with (\ref{eta_conn}) in the
form \cite{Venkataraman:1997xi}
\beq
F_\et(t)\;\equiv\;
\frac{\Nf}{\Nt} C_\et(t)-\frac{\Nf^2}{\Nt^2} D_\et(t)
\label{eta_full}
\eeq
to obtain the full 2-point function of the $\et$ state.
Here $\Nt=2^{d/2}$ denotes the number of ``tastes'' of a staggered field in $d$
spacetime dimensions.
Since $C_\et(t)$ and $F_\et(t)$ fall off exponentially%
%%%
%\footnote{Strictly speaking, there is a transfer matrix argument ensuring this
%only for $\Nf\!=\!2$ and, in a quenched sense, for $\Nf\!=\!0$
%\cite{Bernard:2010qc}. For $\Nf\!=\!1$, there is no such argument (because of
%the rooted determinant), but our data are consistent with $D_\et(t)$ being the
%\emph{difference} of two exponentials for $\Nf\!=\!1$, too. In other words,
%regardless of $\Nf$ the disconnected piece seems to have precisely the form
%needed to make (\ref{eta_full}) a single exponential at $t\!\to\!\infty$.},
%%%
\footnote{Strictly speaking, there is no transfer matrix argument ensuring this
for the case of interest. For $C_\et(t)$ one may be able to construct a
transfer matrix in the partially quenched sense \cite{Bernard:2010qc}. For
$F_\et(t)$, a transfer matrix exits only for $\Nf\!=\!2$. For $\Nf\!=\!1$,
there is no such argument (because of the rooted determinant), but our data are
consistent with $D_\et(t)$ being the \emph{difference} of two exponentials for
$\Nf\!=\!0,1,2$ alike. In other words, regardless of $\Nf$ the disconnected
piece seems to have precisely the form needed to make (\ref{eta_full}) a single
exponential at $t\!\to\!\infty$.},
%%%
at large $t$, with masses $\Met^\mr{conn}$ and $\Met^\mr{full}$ respectively
(where only the latter one is physical), it follows that the ratio of the
disconnected over the connected correlator takes the form
\beq
R_\et(t)\equiv
\frac{D_\et(t)}{C_\et(t)}
\;\longrightarrow\;
\frac{\Nt}{\Nf}-\mr{const}\,
\frac{e^{-\Met^\mr{\;full\;}t}+e^{-\Met^\mr{\;full\;}(T-t)}}
{e^{-\Met^\mr{conn}t}+e^{-\Met^\mr{conn}(T-t)}}
\;\simeq\;
\frac{\Nt}{\Nf}-\mr{const}\,e^{-\Delta\!\Met\,t}
\label{eta_ratio}
\eeq
where $\Delta\!\Met\!\equiv\!\Met^\mr{full}\!-\!\Met^\mr{conn}$, and the
simplification applies for $a\!\ll\!t\!\ll\!T$.
In other words, the prediction is that in 2D the ratio (\ref{eta_ratio}) levels
off at 2 for $\Nf\!=\!1$, and at 1 for $\Nf\!=\!2$.

%%%%%%%%%%%%%%%%%%%%%%%%%%%%%%%%%%%%%%%%%%%%%%%%%%%%%%%%%%%%%%%%%%%%%%%%%%%%%%%

\section{Telling Goldstone bosons from non-Goldstone bosons}

%%%%%%%%%%%%%%%%%%%%%%%%%%%%%%%%%%%%%%%%%%%%%%%%%%%%%%%%%%%%%%%%%%%%%%%%%%%%%%%

\begin{figure}[!tb]
\includegraphics[width=0.49\textwidth]{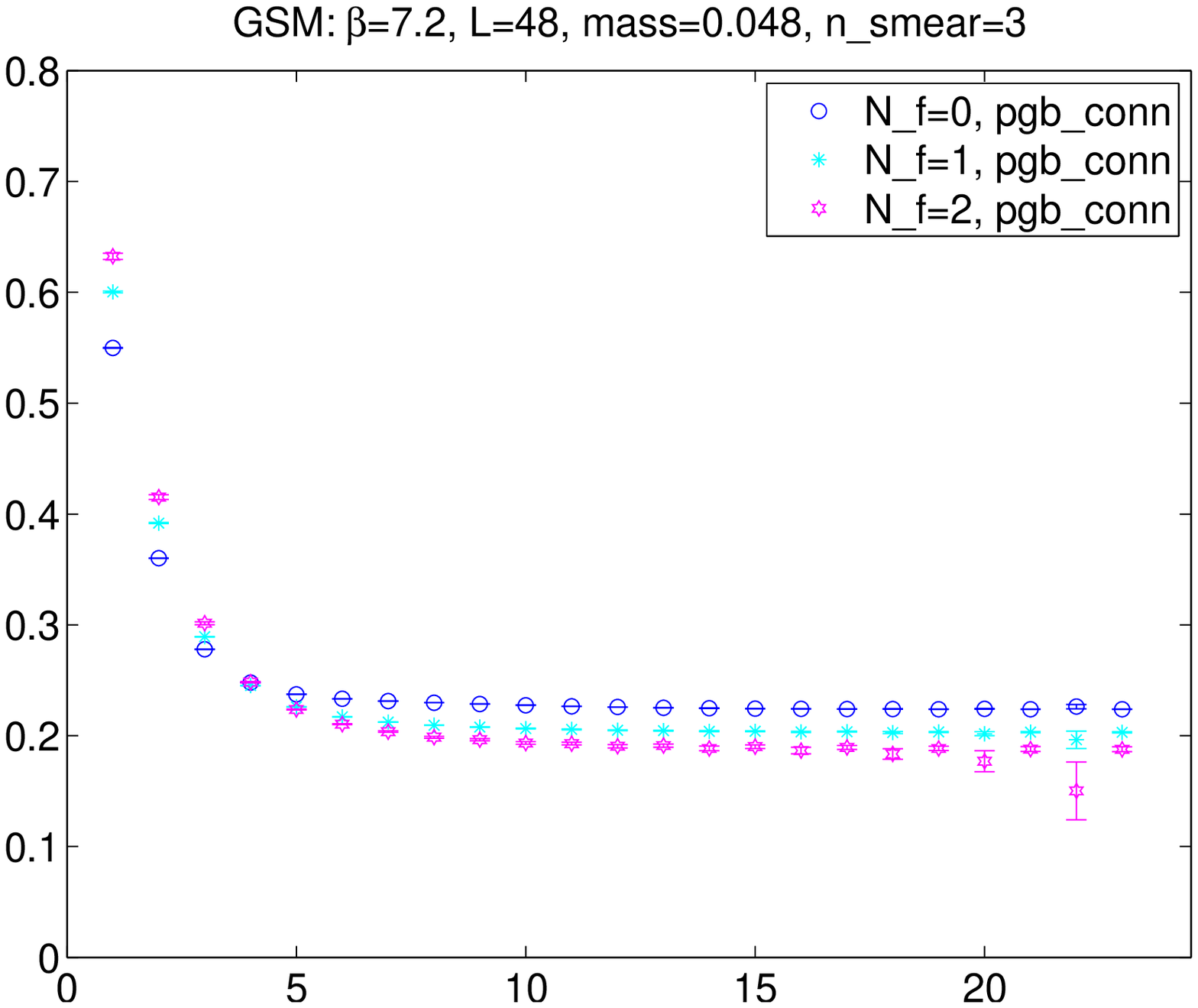}
\includegraphics[width=0.49\textwidth]{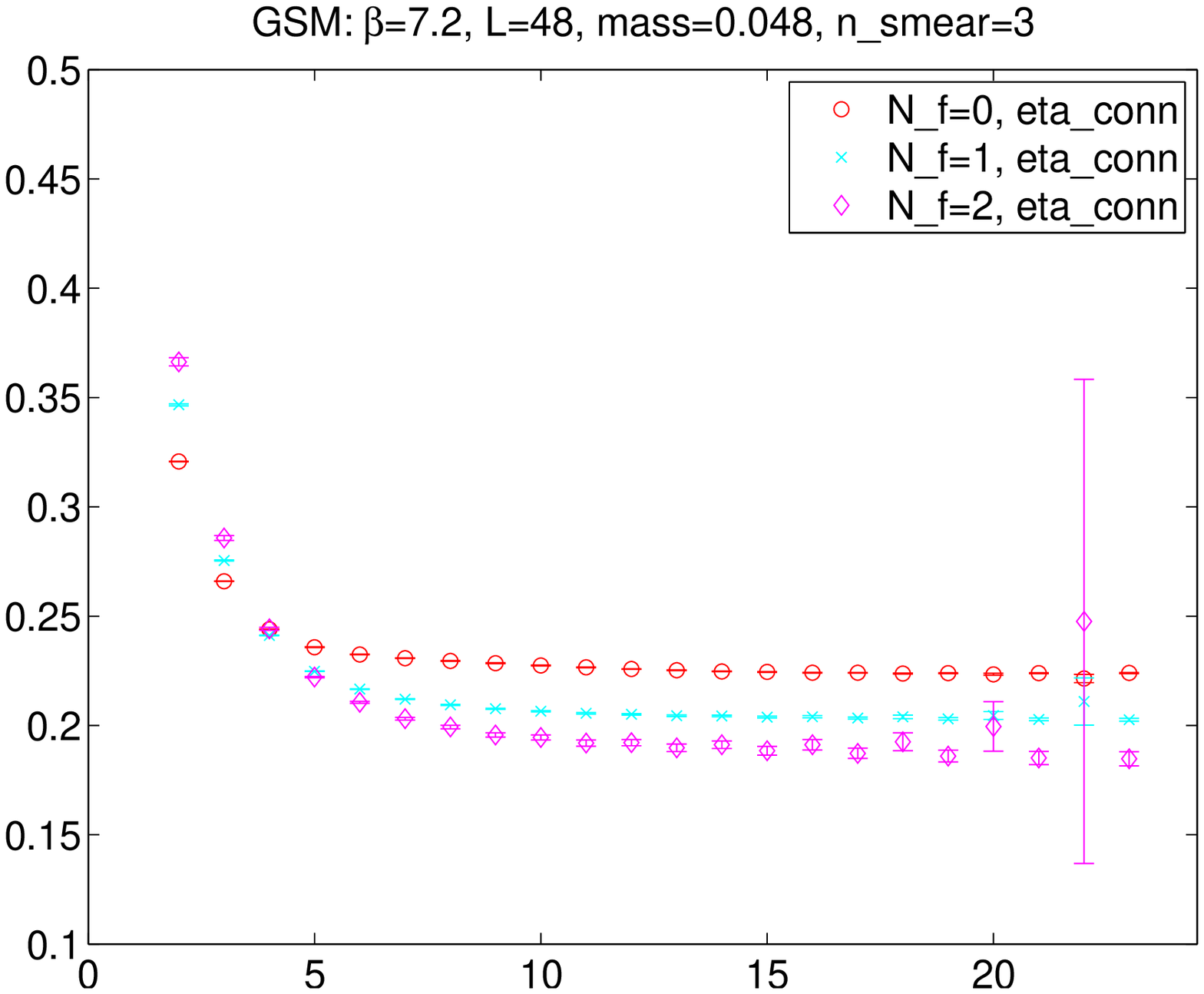}
\caption{\sl\label{fig:all_conn}
$aM_\mr{eff}(t)$ of $C_\pi(t)$ (left) and $C_\et(t)$ (right), at $\be\!=\!7.2$
for the intermediate quark mass.}
\end{figure}

\begin{figure}[!tb]
\includegraphics[width=0.49\textwidth]{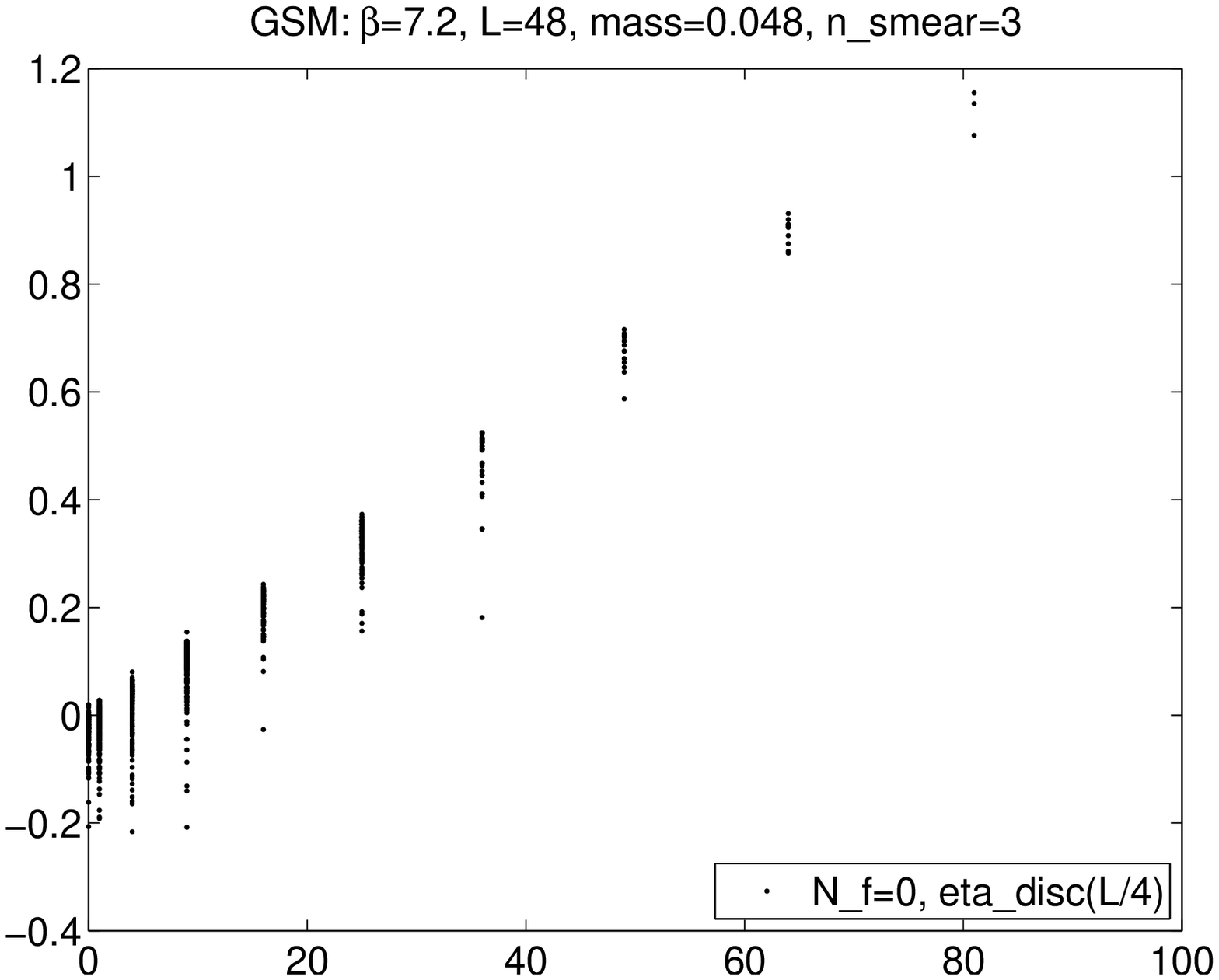}
\includegraphics[width=0.49\textwidth]{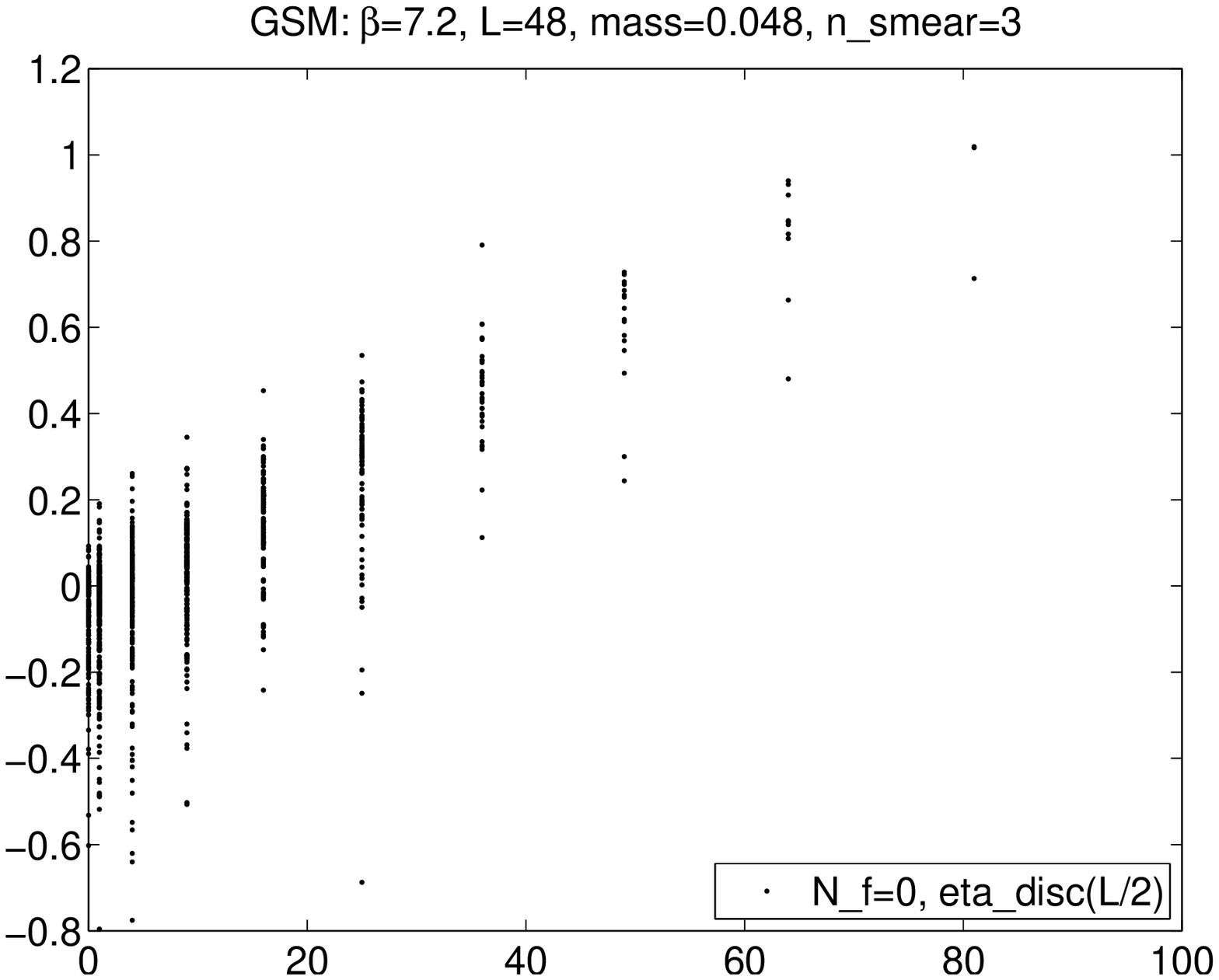}
\caption{\sl\label{fig:top_disc}
Correlation between the disconnected $D_\et(t)$ and the squared topological
charge $q_\mr{geo}^{(3)}$ for $t\!=\!T/4$ (left) and $t\!=\!T/2$ (right).
Data are from the intermediate quark mass run at $\be\!=\!7.2$.}
\end{figure}

\begin{figure}[!tb]
\includegraphics[width=0.49\textwidth]{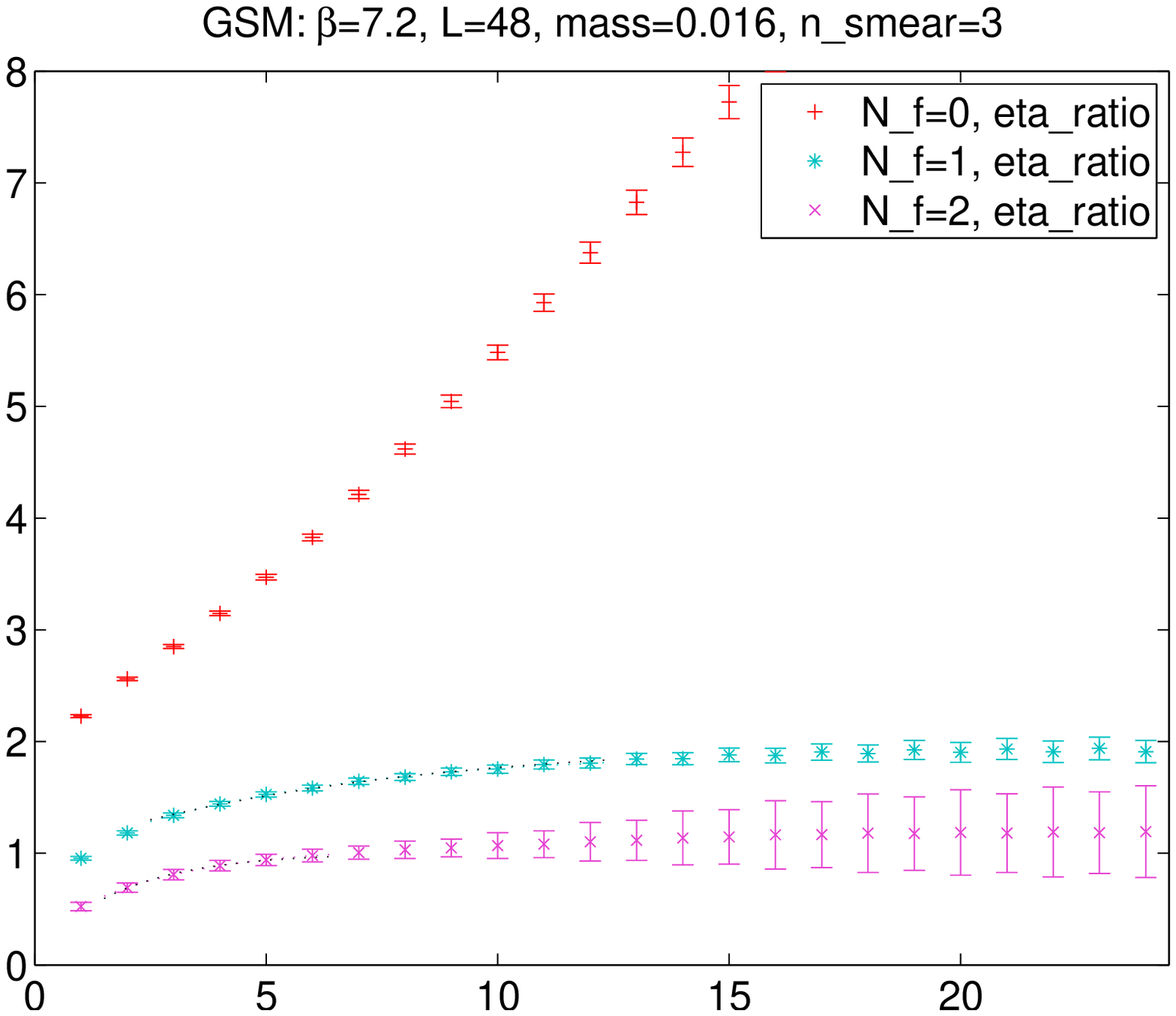}
\includegraphics[width=0.49\textwidth]{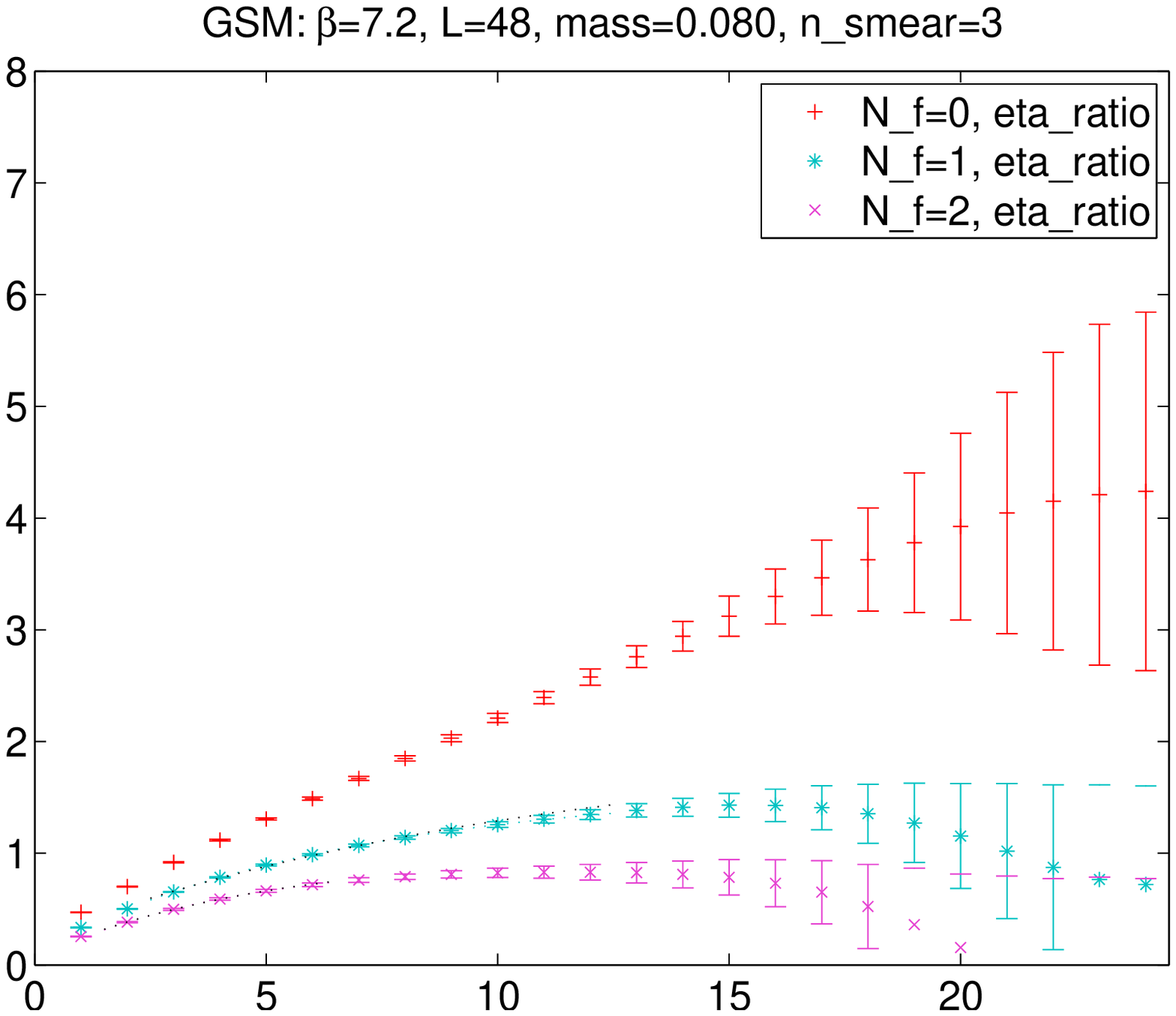}
\caption{\sl\label{fig:eta_ratio}
Ratio $R_\et(t)$ (\ref{eta_ratio}) at $\be\!=\!7.2$ for the smallest (left)
and largest (right) quark mass.}
\end{figure}

With the lattices of Tab.\,\ref{tab:setup} in hand it is advisable to first
check how well the different topological charges agree, how well the overall
distribution is sampled, and whether the disconnected piece $D_\et(t)$ is
indeed sensitive to the overall charge of the background $U$.

The correlation between any pair of the three topological charges considered
is very good; for the larger two $\be$ even the $Z$-factor inherent in
$q_\mr{nai}^\mr{ren}$ and $q_\mr{stag}^\mr{ren}$ is extremely close to 1, as
can be inferred from Fig.\,\ref{fig:topology}.
For $\Nf\!=\!0$ the overall topological charge distribution is nicely sampled
and close to a Gaussian.
Under reweighting to $\Nf\!=\!1$ or $\Nf\!=\!2$ tiny asymmetries seem to get
considerably enhanced.
A feature relevant in what follows is that the dynamical distributions are
\emph{narrower} than the quenched one; with observables which are sensitive to
topology it is useful to have the tails ``oversampled'' and to reduce their
weight in the analysis.

After these checks have been carried out successfully, it is straightforward to
determine the pion mass $a\Mpi$ for all couplings and quark masses.
To this aim we consider the effective mass
\beq
a\Mpi(t)=\frac{1}{2}\log(
[C(t\!-\!1)+\sqrt{C(t\!-\!1)^2-C(T/2)^2}]/
[C(t\!+\!1)+\sqrt{C(t\!+\!1)^2-C(T/2)^2}])
\eeq
with $C\!\equiv\!C_\pi$, which is designed to compensate for the influence of
the periodic boundary conditions.
A typical example is shown in Fig.\,\ref{fig:all_conn}.
A nice plateau is observed; albeit with tiny oscillations%
\footnote{This may signal the presence of a parity partner
\cite{Bazavov:2009bb} on odd timeslices. In our analysis we use
$\Mpi(T/2\!-\!1)$ which, in turn, is based on the correlator $C_\pi(t)$ on the
even timeslices $t=T/2\!-\!2,T/2$.}
which grow towards the center of the box.
Under reweighting to $\Nf\!=\!1$ or $\Nf\!=\!2$ this effect gets enhanced.
For the connected piece of the eta (which is an unphysical state) similar
results are obtained.
Throughout, the difference between these two masses is tiny.

Turning to the disconnected piece $D_\et(t)$, we first consider its correlation
with the topological charge of the background.
Example results are presented in Fig.\,\ref{fig:top_disc}.
For these $t$ the sub-ensemble average $\<D_\et(t)\>_{q}$ seems to be a linear
function of $q^2$ (which, from a glimpse at (\ref{eta_disc}) and the definition
of $q_\mr{stag}[U]$ is plausible).
The issue most relevant is whether reasonable results for the
disconnected-over-connected ratio (\ref{eta_ratio}) are obtained.
Typical results are presented in Fig.\,\ref{fig:eta_ratio}.
We obtain a rather clear signal up to about $t\!=\!T/4$ and find a
qualitatively different behavior for the three $\Nf$ shown.
In the quenched case the pattern is consistent with a linear rise (with a slope
which clearly depends on the quark mass).
After reweighting to $\Nf\!=\!1$ or $\Nf\!=\!2$ the behavior is consistent with
the prediction (\ref{eta_ratio}).
A typical problem with the disconnected piece is that at large $t$ the data may
go astray without the statistical error giving a hint of this (we tried
several jackknife blocksizes).
Still, we can extract the mass gap $a\Delta\!\Met$ from a fit to
(\ref{eta_ratio}) at intermediate $t$, i.e.\ before the noise prevails.
The results such obtained are collected in Tab.\,\ref{tab:results}, along with
the $a\Mpi$ values determined previously.

\begin{table}[!tb]
%%% ! grep -v # results_01p8_24.dat | awk '/./' > tmp.dat
%%% ! grep -v # results_03p2_32.dat | awk '/./' > tmp.dat
%%% ! grep -v # results_07p2_48.dat | awk '/./' > tmp.dat
%%% ! grep -v # results_12p8_64.dat | awk '/./' > tmp.dat
%%% data=load('tmp.dat'); ! rm -f tmp.dat;
%%% data=permute(reshape(data,[3,5,6]),[1,3,2]);
%%% data(4,[1:2:5],:)=data(2,[1:2:5],:)+data(3,[1:2:5],:);
%%% data(4,[2:2:6],:)=sqrt(data(2,[2:2:6],:).^2+data(3,[2:2:6],:).^2);
%%% format short; round(1000*squeeze(data(1,:,:)))/1000 %%% slot_1 is Mpi
%%% format short; round(1000*squeeze(data(2,:,:)))/1000 %%% slot_2 is Met_conn
%%% format short; round( 100*squeeze(data(3,:,:)))/ 100 %%% slot_3 is Met_delt
%%% format short; round( 100*squeeze(data(4,:,:)))/ 100 %%% slot_4 is Met_full
\centering
\footnotesize
\begin{tabular}{|@{\,}c@{\,}c@{\,}c@{\,}|
 @{\,}c@{\,\,}c@{\,\,}c@{\,\,}c@{\,\,}c@{\,}|
 @{\,}c@{\,\,}c@{\,\,}c@{\,\,}c@{\,\,}c@{\,}|}
\hline
 & & & & & $a\Mpi$ & & & & & $a\Delta\!\Met$ & & \\
\hline
$\be$ & $L/a$ & $\Nf$ &
$m_1$ & $m_2$ & $m_3$ & $m_4$ & $m_5$ &
$m_1$ & $m_2$ & $m_3$ & $m_4$ & $m_5$ \\
\hline
 1.8 & 24 & 0 & 0.279(01) & 0.368(01) & 0.444(01) & 0.513(01) & 0.579(01) &    ---   &    ---   &    ---   &    ---   &    ---   \\
     &    & 1 & 0.220(01) & 0.318(01) & 0.402(01) & 0.478(01) & 0.550(01) & 0.39(02) & 0.36(07) & 0.33(03) & 0.31(04) & 0.30(05) \\
     &    & 2 & 0.199(06) & 0.292(02) & 0.374(02) & 0.454(01) & 0.530(01) & 0.83(42) & 1.05(60) & 0.63(08) & 0.57(14) & 0.53(21) \\
 3.2 & 32 & 0 & 0.215(01) & 0.281(01) & 0.337(01) & 0.389(01) & 0.439(01) &    ---   &    ---   &    ---   &    ---   &    ---   \\
     &    & 1 & 0.166(01) & 0.240(01) & 0.304(01) & 0.362(01) & 0.417(01) & 0.28(02) & 0.22(01) & 0.21(01) & 0.21(01) & 0.18(01) \\
     &    & 2 & 0.154(07) & 0.217(04) & 0.283(01) & 0.342(01) & 0.400(01) & 1.41(97) & 0.48(02) & 0.39(06) & 0.36(02) & 0.40(06) \\
 7.2 & 48 & 0 & 0.142(01) & 0.186(01) & 0.224(01) & 0.260(01) & 0.294(01) &    ---   &    ---   &    ---   &    ---   &    ---   \\
     &    & 1 & 0.111(02) & 0.159(01) & 0.203(01) & 0.242(01) & 0.279(01) & 0.17(02) & 0.14(01) & 0.12(01) & 0.11(01) & 0.12(03) \\
     &    & 2 & 0.091(16) & 0.137(06) & 0.188(02) & 0.230(01) & 0.269(01) & 0.45(07) & 0.29(02) & 0.28(01) & 0.24(02) & 0.22(02) \\
12.8 & 64 & 0 & 0.107(01) & 0.140(01) & 0.169(01) & 0.196(01) & 0.221(01) &    ---   &    ---   &    ---   &    ---   &    ---   \\
     &    & 1 & 0.087(02) & 0.123(01) & 0.153(01) & 0.183(01) & 0.210(01) & 0.14(02) & 0.12(01) & 0.10(02) & 0.09(01) & 0.09(02) \\
     &    & 2 & 0.086(07) & 0.114(02) & 0.143(02) & 0.175(01) & 0.201(02) & 0.32(05) & 0.24(04) & 0.23(02) & 0.19(01) & 0.17(02) \\
\hline
 3.2 & 24 & 0 & 0.231(01) & 0.285(01) & 0.338(01) & 0.390(01) & 0.439(01) &    ---   &    ---   &    ---   &    ---   &    ---   \\
     &    & 1 & 0.182(01) & 0.246(01) & 0.306(01) & 0.363(01) & 0.416(01) & 0.26(01) & 0.23(01) & 0.21(01) & 0.19(01) & 0.18(01) \\
     &    & 2 & 0.173(02) & 0.229(01) & 0.288(01) & 0.346(01) & 0.400(01) &2.12(1.56)& 0.64(16) & 0.47(02) & 0.37(04) & 0.33(05) \\
 3.2 & 40 & 0 & 0.214(01) & 0.280(01) & 0.337(01) & 0.389(01) & 0.439(01) &    ---   &    ---   &    ---   &    ---   &    ---   \\
     &    & 1 & 0.158(02) & 0.241(01) & 0.302(01) & 0.363(01) & 0.417(01) & 0.28(13) & 0.26(03) & 0.21(02) & 0.18(03) & 0.21(04) \\
     &    & 2 & 0.138(04) & 0.220(04) & 0.276(05) & 0.347(02) & 0.402(01) &2.75(2.72)& 1.23(83) & 0.41(25) & 0.38(12) & 0.36(04) \\
\hline
\end{tabular}
\caption{\sl\label{tab:results}
Measured $a\Mpi$ (Goldstone boson) and $a\Delta\!\Met$ (excess of the
taste-and-flavor singlet pseudoscalar state) for all $\be$, $L/a$, $\Nf$, and
$m$. The quark masses are given in Tab.\,1.}
\end{table}

The lower part of Tab.\,\ref{tab:results} contains the results of a
finite-volume scaling study at $\be\!=\!3.2$ (since finite volume effects
relate to infrared physics the restriction to one $\be$ is permissible).
It seems that in the extra-small volume the pion mass is affected for the
lightest two quark mass values.
In the standard volume only the lightest quark mass data suffer from small
finite volume artefacts.
In case of the mass gap $a\Delta\!\Met$ no finite volume effects are
found for $\Nf\!=\!1$, while for $\Nf\!=\!2$ the quality of the data is
less convincing, which likely indicates the limitations of the reweighting
method (fortunately, we are mostly interested in $a\Delta\!\Met$ for
$\Nf\!=\!1$).

\begin{figure}[!tb]
\includegraphics[width=0.49\textwidth]{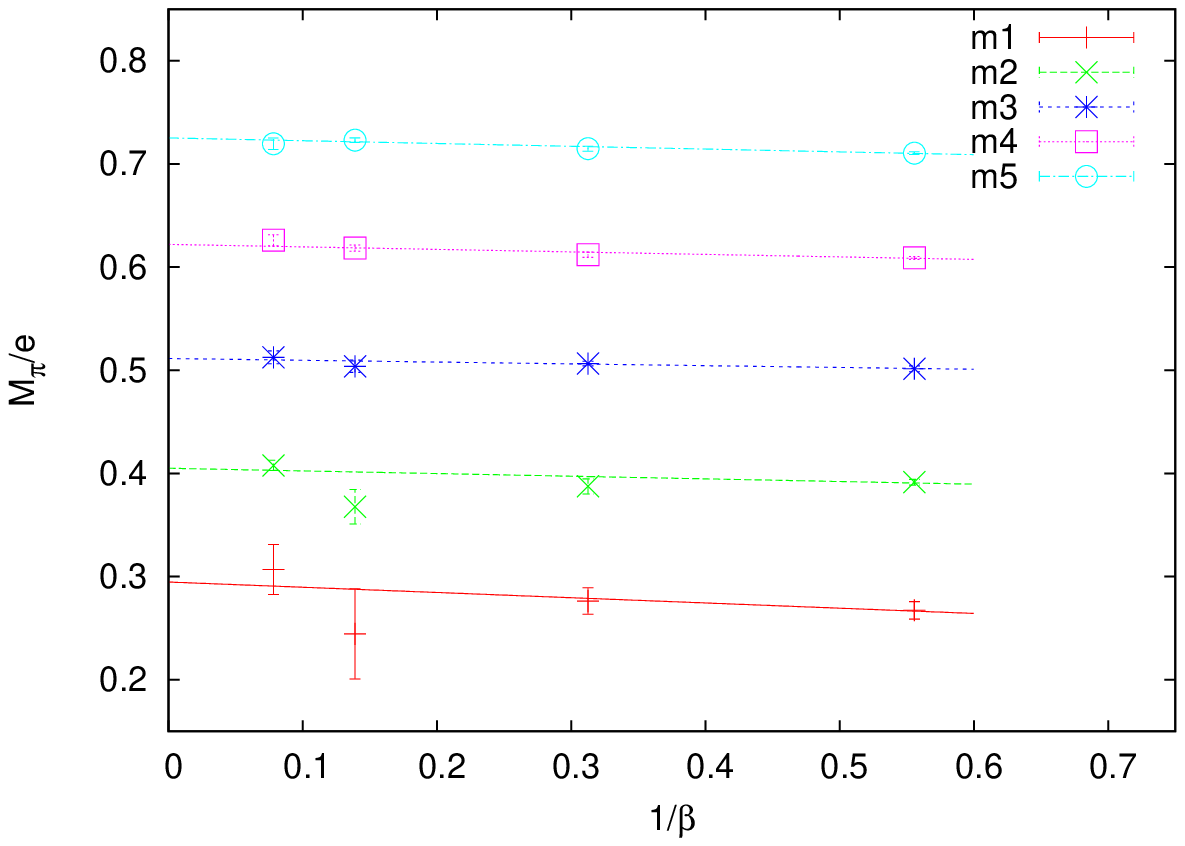}
\includegraphics[width=0.49\textwidth]{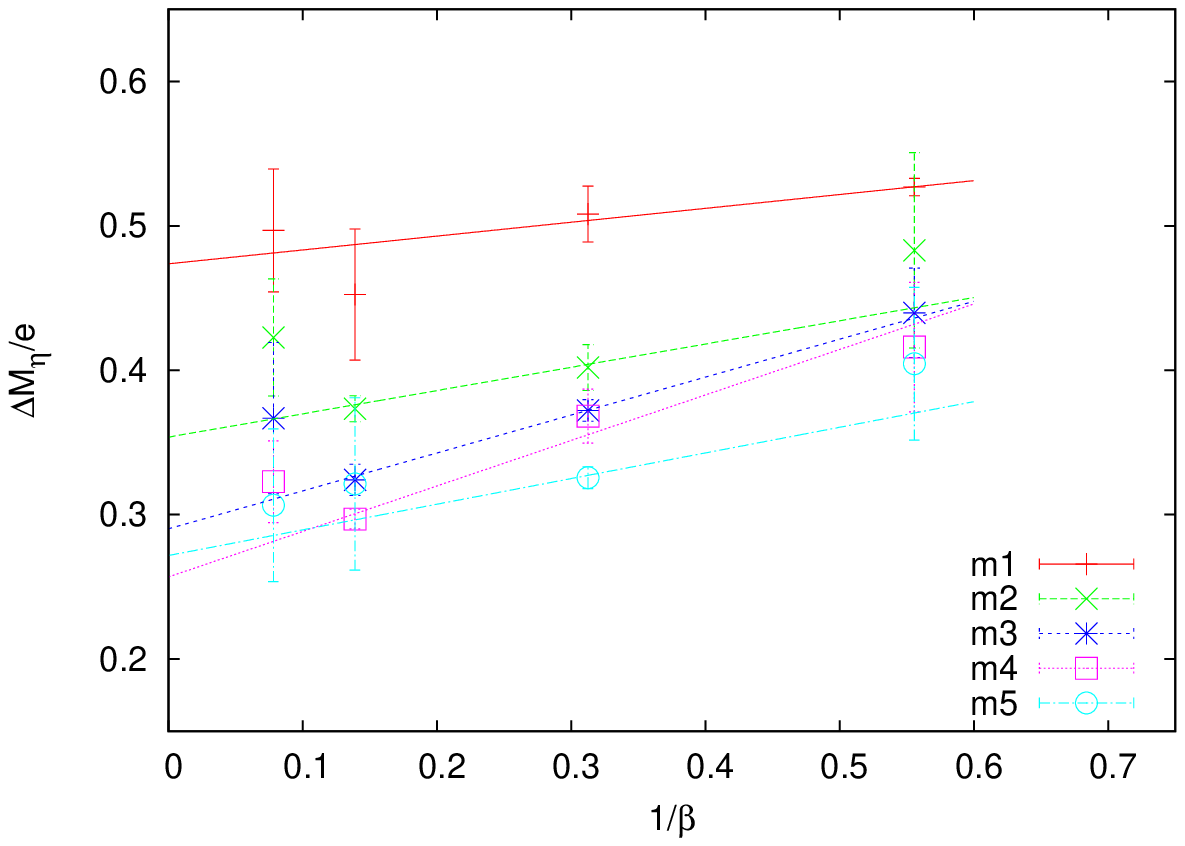}
\vspace*{-4mm}
\caption{\sl\label{fig:cont}
Continuum extrapolation of $\Mpi/e$ ($\Nf\!=\!2$, left) and $\Delta\!\Met/e$
($\Nf\!=\!1$, right) versus $a^2$.}
\end{figure}

\begin{figure}[!tb]
\includegraphics[width=0.49\textwidth]{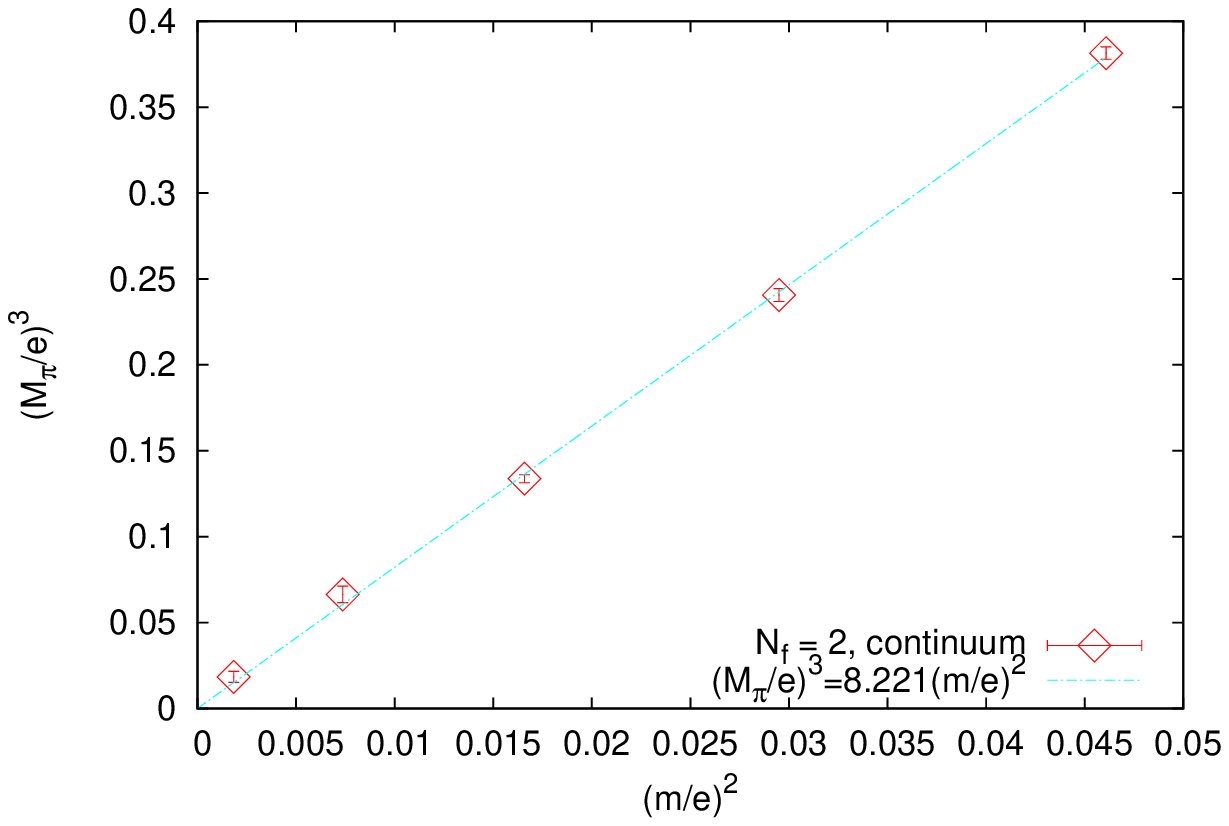}
\includegraphics[width=0.49\textwidth]{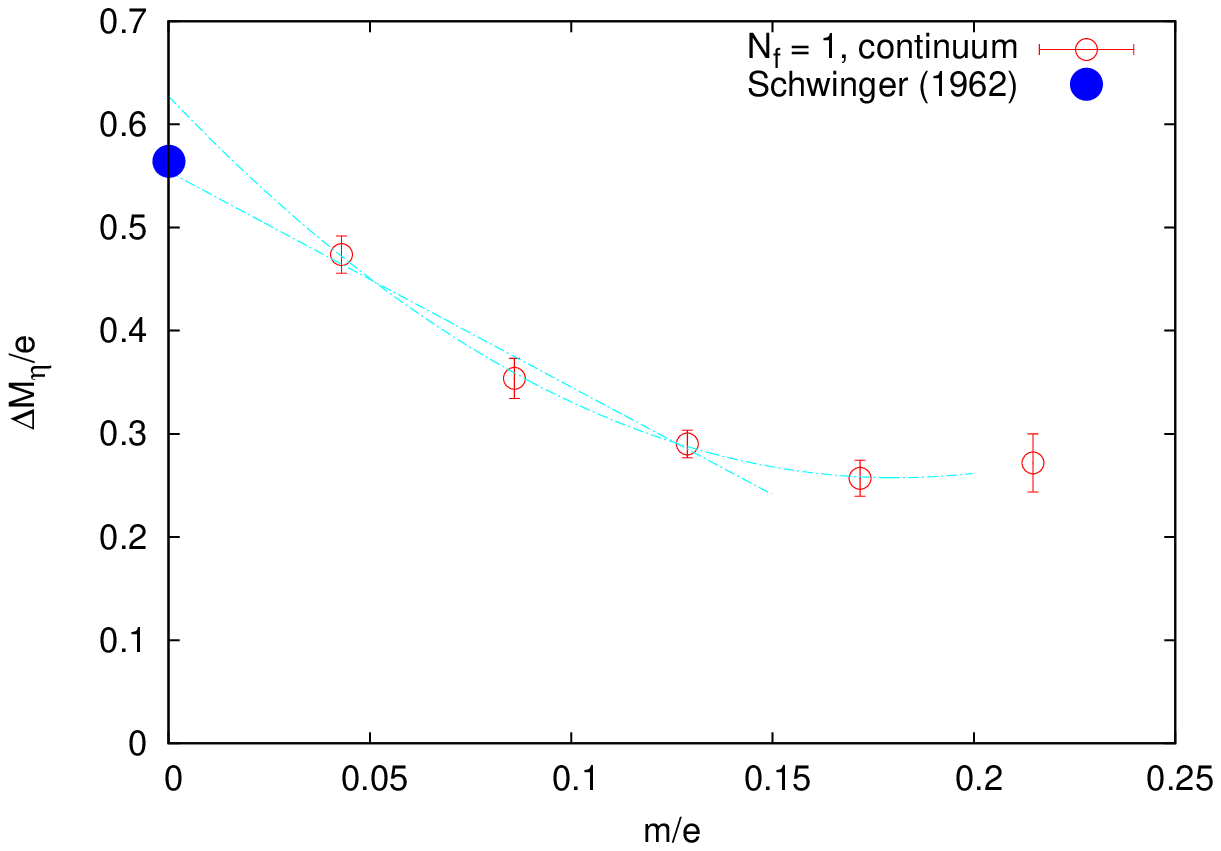}
\vspace*{-4mm}
\caption{\sl\label{fig:chir}
Chiral extrapolation of $(\Mpi/e)^3$ versus $(m/e)^2$ (left) and
$\Delta\!\Met/e$ versus $m/e$ (right).}
\end{figure}

The next step is a continuum extrapolation of the meson masses obtained.
We do this both for $\Mpi/e$ in the $\Nf\!=\!2$ theory (to test the prediction
by Smilga \cite{Smilga:1996pi}, see above) and for the main object of interest
$\Delta\!\Met/e$ in the $\Nf\!=\!1$ theory.
As mentioned in Sec.\,2 the simulations were carried out at fixed physical
quark masses (i.e.\ $m/e$), such that the data of Tab.\,\ref{tab:results} can
be continuum extrapolated without interpolation in $m$.
These extrapolations are shown in Fig.\,\ref{fig:cont}.
It seems that all four lattice spacings are in the Symanzik scaling regime;
we obtain acceptable linear fits with two degrees of freedom.

With these continuum results for $\Mpi/e$ (for $\Nf\!=\!2$) and
$\Delta\!\Met/e$ (for $\Nf\!=\!1$) in hand, it is interesting to consider their
quark mass dependence.
In the left panel of Fig.\,\ref{fig:chir} we plot $(\Mpi/e)^3$ as a function of
$(m/e)^2$.
Here, the lightest pion mass has been adjusted by the finite volume correction
factor $0.138/0.154\!=\!0.896$ (taken from Tab.\,\ref{tab:results}).
Using a 1-parameter ansatz fits the data with $\ch^2/\mr{d.o.f}\!=\!1.13$ and
a slope parameter of $8.221(63)$.
This is in reasonable agreement with Smilga's prediction that this parameter
should be $2.008^3\!=\!8.096$ \cite{Smilga:1996pi}.

Last but not least let us consider the quark mass dependence of the mass gap
$\Delta\!\Met/e$ in the $\Nf\!=\!1$ theory, as shown in the right panel of
Fig.\,\ref{fig:chir}.
Being unaware of an analytic prediction, it is not entirely clear which powers
of $\Delta\!\Met/e$ and $m/e$ one should choose to display the data, and we
opt for staying without additional powers.
Fortunately, in this representation the quark mass dependence seems mild, and
both a linear and a quadratic fit with one degree of freedom describe the data
convincingly.
%%% linear: 0.554134 +/- 0.03334
%%% quadra: 0.627198 +/- 0.01659
Taking half of the spread as the systematic error, these fits predict
$\Delta\!\Met/e=0.591(25)(37)$ in the chiral limit%
\footnote{Here we assume that the unphysical mass $\Met^\mr{conn}$ that
belongs to $C_\et(t)$ vanishes in the combined continuum and chiral limit.
Given that our data are consistent with $a\Mpi\!-\!a\Met^\mr{conn}$ being a
cut-off effect (cf.\ Fig.\,\ref{fig:all_conn}), this seems to be the case.
The alternative of extrapolating the physical $\Met$ seems more susceptible to
the choice of the powers on $\Met/e$ and $m/e$, which results in a larger
uncertainty of the extrapolated result.},
which is in perfect agreement with Schwinger's analytic result
$\Met/e=\Delta\!\Met/e=1/\sqrt{\pi}=0.56419$ \cite{Schwinger:1962tp}.

%%%%%%%%%%%%%%%%%%%%%%%%%%%%%%%%%%%%%%%%%%%%%%%%%%%%%%%%%%%%%%%%%%%%%%%%%%%%%%%

\section{Summary}

%%%%%%%%%%%%%%%%%%%%%%%%%%%%%%%%%%%%%%%%%%%%%%%%%%%%%%%%%%%%%%%%%%%%%%%%%%%%%%%

Triggered by the criticism of \cite{Jansen:2003nt,Creutz:2007yg,Creutz:2007rk}
the validity of the ``rooting trick'' in studies with staggered quarks has been
the subject of an intense debate in the lattice community.
Several review talks at major conferences
\cite{Durr:2005ax,Sharpe:2006re,Kronfeld:2007ek,Golterman:2008gt} presented a
wealth of numerical and analytical evidence in favor of the procedure, but so
far the ``experimentum crucis'', i.e.\ a direct test of $\et'$-phenomenology in
QCD with rooted staggered quarks remained elusive, due to noise issues
\cite{Venkataraman:1997xi,Gregory:2007ev,Gregory:2008mn,Gregory:2011sg}.

This paper is based on the observation that there is no strict need to
investigate the topic in QCD, since the conceptual issue is one-to-one
matched in the massive Schwinger model with 1 flavor, which is much simpler
to simulate.
We demonstrated that in this case it is possible to obtain conclusive results
for the disconnected-over-connected ratio (\ref{eta_ratio}).
As shown in Fig.\,\ref{fig:eta_ratio} the prediction (\ref{eta_ratio}) with
$\Nf\!=\!1$ and $\Nt\!=\!2$ is beautifully confirmed.
In consequence the mass gap $\Delta\!\Met$ or the physical mass $\Met$ in the
1-flavor theory can be determined with sufficient precision, so that a combined
continuum and chiral extrapolation is possible.
The result is in perfect agreement with the analytical prediction
$\Met\!=\!e/\sqrt{\pi}$ by Schwinger \cite{Schwinger:1962tp}.

It seems this is the first \emph{ad oculos} demonstration that the staggered
setup --~with the rooting trick in the functional measure~-- treats the
contribution of the axial anomaly to the particle spectrum correctly.
With this result, and in view of
\cite{Venkataraman:1997xi,Gregory:2007ev,Gregory:2008mn,Gregory:2011sg}, one
may predict that the outcome of a similar study in QCD will be identical, once
sufficient CPU power is available.

\bigskip

\noindent{\bf Acknowledgments}:
%%%
%...
%%%
It is a pleasure to acknowledge useful correspondence with Steve Sharpe,
Claude Bernard and Zolt\'an Fodor.
Partial support was provided by Bern University, DESY Zeuthen, and through
SFB/TR-55.
Computations were performed on a stand-alone PC at JSC.
%%%

%%%%%%%%%%%%%%%%%%%%%%%%%%%%%%%%%%%%%%%%%%%%%%%%%%%%%%%%%%%%%%%%%%%%%%%%%%%%%%%

%%%%%%%%%%%%%%%%%%%%%%%%%%%%%%%%%%%%%%%%%%%%%%%%%%%%%%%%%%%%%%%%%%%%%%%%%%%%%%%

\end{document}